

\magnification\magstep1 
\baselineskip = 0.6 true cm
\parskip=0.5 true cm
                           
  \def\sa{\vskip 0.30 true cm}
  \def\sb{\vskip 0.60 true cm}
  
  \nopagenumbers 
       
  \hsize = 17.5 true cm
  \vsize = 25.2   true cm



\rightline{\bf LYCEN 9117}
\rightline{05 July 1991}

\sa
\sb
\sa

\noindent {\bf SYMMETRY ADAPTATION AND TWO-PHOTON SPECTROSCOPY}

\noindent {\bf OF IONS IN MOLECULAR OR SOLID--STATE FINITE SYMMETRY$^*$}

\sa
\sb

\noindent M.~Kibler and M.~Daoud 

\sa

\noindent Institut de Physique Nucl\'eaire de Lyon \hfill\break
IN2P3-CNRS et Universit\'e Claude Bernard \hfill\break
F-69622 Villeurbanne Cedex, France 

\sa
\sa
\sb
\sb

\baselineskip = 0.5 true cm

\sa

\noindent {\bf Abstract}. Finite symmetry adaptation techniques are 
applied to the determination of the intensity strength of
two-photon transitions for ions with one partly-filled shell 
$n \ell$ in crystalline environments of symmetry $G$. We 
treat the case of intra-configurational ($n \ell^N \to n \ell^N)$
transitions as well as the case of inter-configurational 
($n \ell^N \to n \ell^{N-1} n' \ell'$ with 
$(-)^{\ell + \ell'} = - 1$) transitions. In both cases, the 
Wigner-Racah algebra of the chain $O(3) \supset G$ allows to 
extract the polarization dependence from the intensity. The 
reported results are valid for any strength of the crystalline 
field.  

\sa
\sb
\sa
\sb
\sa
\sb
\sb
\sb

\noindent $^*$ Invited paper at the 
``V Workshop on Symmetry Methods in Physics'', 
Obninsk, Kaluga region, USSR, 8~-~12 July 1991. 
Published in {\bf Symmetry Methods in Physics}, 
Eds. Yu.F.~Smirnov and R.M.~Asherova 
(Institute of Physics and Power Engineering, Obninsk, 1992), 
(p. 37-49). 

\vfill\eject
\baselineskip = 0.65 true cm
\vglue 2.5 true cm

\noindent {\bf SYMMETRY ADAPTATION AND TWO-PHOTON SPECTROSCOPY}

\noindent {\bf OF IONS IN MOLECULAR OR SOLID--STATE FINITE SYMMETRY}

\sa
\sb

\noindent M.~Kibler and M.~Daoud 

\sa

\noindent Institut de Physique Nucl\'eaire de Lyon \hfill\break
IN2P3-CNRS et Universit\'e Claude Bernard \hfill\break
F-69622 Villeurbanne Cedex, France 
\sa
\sa
\sb
\sb
\baselineskip = 0.5 true cm
\sa
\noindent {\bf Abstract}. Finite symmetry adaptation techniques are 
applied to the determination of the intensity strength of
two-photon transitions for ions with one partly-filled shell 
$n \ell$ in crystalline environments of symmetry $G$. We 
treat the case of intra-configurational ($n \ell^N \to n \ell^N)$
transitions as well as the case of inter-configurational 
($n \ell^N \to n \ell^{N-1} n' \ell'$ with 
$(-)^{\ell + \ell'} = - 1$) transitions. In both cases, the 
Wigner-Racah algebra of the chain $O(3) \supset G$ allows to 
extract the polarization dependence from the intensity. The 
reported results are valid for any strength of the crystalline 
field.  
\sa
\sb
\sa
\sb
\baselineskip = 0.65 true cm
\noindent {\bf  I. INTRODUCTION}

Symmetry adaptation techniques, based on the use of chains of 
groups $O(3) \supset G$, have been developed by many authors in 
the last twenty years. In particular, numerous studies have 
been achieved in connection with crystal- and ligand-field 
theories (see Refs.~[1-3] and references therein).

The aim of this paper is to show how Wigner-Racah calculus 
for a chain of type $O(3) \supset G$ (in terms of simple or 
double groups) can be applied to the determination of the 
intensity of two-photon transitions for an ion with 
configuration $n \ell^N$ in a molecular or solid-state 
environment with symmetry $G$. (For classification and 
symmetry-breaking purposes, the group $G$ may be replaced by a 
chain of subgroups of $O(3)$, the relevant symmetry group being 
one of the groups of the chain.)  

The r\^ole of symmetries in two-photon spectroscopy of 
partly-filled shell ions in finite symmetry, of interest in 
molecular and solid-state physics, is touched upon in 
Refs.~[4-8]. In Ref.~[7], the accent is put on the transition 
matrix elements between initial and final state vectors while 
emphasize is on the intensity strength in Ref.~[8] and in the 
present paper. 

Two distinct cases are studied in this work. The case of 
($n \ell^N \to n \ell^N$) intra-configurational two-photon 
transitions, which are parity allowed, is worked out in section II 
and the one of 
($n \ell^N \to n \ell^{N-1} n' \ell'$ with $(-)^{\ell + \ell'}=-1$) 
inter-configurational two-photon transitions, which are parity forbidden, is 
examined in section III.  

\noindent {\bf II. INTRA-CONFIGURATIONAL TWO-PHOTON TRANSITIONS}

\noindent {\bf Preliminaries}

We know that the electronic transition matrix element $M_{i \to f}$ between an
initial state $i$ and a final state $f$ is, in the framework of the dipolar
approximation, given by
$$
M_{i \to f} \; = \; \sum_v \; {{1}\over {\Delta_1}} \; 
\left(f \vert \vec D. \, \vec {\cal E}_2 \vert v \right) 
\left(v \vert \vec D. \, \vec {\cal E}_1 \vert i \right) 
                  + \sum_v \; {{1}\over {\Delta_2}} \; 
\left(f \vert \vec D. \, \vec {\cal E}_1 \vert v \right) 
\left(v \vert \vec D. \, \vec {\cal E}_2 \vert i \right) 
\eqno (1)
$$
The two summations in (1) have to be extended over all the 
(virtual) intermediate states $v$ having a parity different 
from the one of the states $i$ and $f$. Furthermore, we have
$$
\Delta_\lambda \; = \; \hbar \omega_{\lambda} \; - \; E_v 
\qquad \hbox {for} \qquad \lambda = 1, 2
\eqno (2)
$$
where $E_v$ is the energy of the state $v$ with respect to 
that of the state $i$ and
$\hbar \omega_{\lambda}$ the energy of the photon no.~$\lambda$. (For 
Raman scattering, the sign of $\hbar \omega_2$ has to be 
changed.) In equation (1), the quantity 
$\vec D . \, {\vec {\cal E}}_{\lambda}$ (with $\lambda = 1,2$) stands for 
the scalar product of the electric dipolar moment operator  
$$
\vec D = - e \; \sum^N_{j = 1} \; \vec r_j \eqno (3)
$$
for the $N$ electrons and the unit polarization vector 
${\vec {\cal E}}_\lambda$ for the photon no.~$\lambda$. (We use 
single mode excitations, of polarization 
${\vec {\cal E}}_{\lambda}$ and energy $\hbar \omega_{\lambda}$, 
of the radiation field.) The two photon beams 
can be polarized either circularly with
$$
({\cal E}_\lambda)_q = - \delta (q,-1) 
\quad \hbox {if} \quad {\vec {\cal E}}_\lambda = {\vec e}_{+1} 
\quad \hbox {while} \quad 
({\cal E}_\lambda)_q = - \delta (q,+1) 
\quad \hbox {if} \quad {\vec {\cal E}}_\lambda = {\vec e}_{-1} 
\quad \hbox {for} \quad \lambda = 1,2
\eqno (4)
$$
or linearly with
$$
({\cal E}_\lambda)_0 = \cos \theta_\lambda \qquad 
({\cal E}_\lambda)_{\pm 1} = \mp
{1 \over {\sqrt 2}} \; \sin \theta_\lambda \; \exp (\pm i \varphi_\lambda) 
\quad \hbox {for} \quad \lambda = 1,2 
\eqno (5)
$$
In equations (4) and (5), we use the components
$$
({\cal E}_\lambda)_q = 
  {\vec {\cal E}}_\lambda . \, {\vec e}_q 
  \qquad \hbox {for} \qquad \lambda = 1,2 
  \qquad \hbox {and} \qquad q = - 1, 0, 1 
\eqno (6)
$$
in the standard spherical basis 
$\left(\vec e_{-1}, \vec e_0, \vec e_{+1}\right)$.
In the case of a linear polarization, the angles 
$(\theta_\lambda, {\varphi}_\lambda)$ 
are the polar angles of the polarization vector 
${\vec {\cal E}}_\lambda$ 
($\lambda = 1,2$) with respect to the crystallographic axis. 
For two-photon absorption, only one sum occurs in (1) when the 
two photons are identical. 

Equation (1) can be derived from the time-dependent perturbation theory 
[9,10] and goes back to the work of G\"oppert-Mayer. It
is also possible to derive it, in an elegant way, from the method of the
resolvent operator [10,11].

\noindent {\bf State vectors}

The initial state $i$ with symmetry $\Gamma$ is characterized by the state
vectors $\vert i \Gamma \gamma)$ where $\gamma$ 
($\gamma = 1$, 2, $\cdots$, dim~$\Gamma$) is a
multiplicity label to be used if the dimension dim~$\Gamma$ of the irreducible
representation class (IRC) $\Gamma$ of the group $G$ is greater than 1. The
state vector $\vert i \Gamma \gamma)$ is taken in the form
$$
\vert i \Gamma \gamma) \equiv \vert n \ell^N  i \Gamma \gamma) = 
\sum_{\alpha S L J a} \; \vert n \ell^N \alpha S L J a \Gamma \gamma) 
\; c(\alpha S L J a \Gamma; i) \eqno (7)
$$
in terms of the $O(3) \supset G$ symmetry adapted state vectors
$$
\vert n \ell^N \alpha S L J a \Gamma \gamma) \; = \; \sum_{M=-J}^J \; 
\vert n \ell^N \alpha S L J M) \;  (J M \vert J a \Gamma \gamma) \eqno (8)
$$
The coefficients $(JM \vert J a \Gamma \gamma)$ in (8) are reduction
coefficients to pass from               the chain $O(3) \supset O(2)$ 
characterizing the $\{ JM \}$ scheme to the chain $O(3) \supset G$ 
characterizing the $\{ J a \Gamma \gamma \}$ scheme ; they 
depend on the group $G$ with a certain degree of freedom 
emphasized by the branching multiplicity label $a$ to be used when $\Gamma$
occurs several times in the IRC $(J)$ of $O(3)$. 
In contradistinction, the coefficients $c(\alpha S L J a \Gamma ; i)$ 
in (7) depend on the Hamiltonian employed for obtaining
the initial state $i$. Similarly, for the final state $f$ with symmetry 
$\Gamma'$, we have the state vectors 
$$
\vert f \Gamma' \gamma') \; \equiv \; \vert n \ell^N f \Gamma' \gamma') 
\; = \; \sum_{\alpha' S' L' J' a'} \; \vert n \ell^N \alpha' S' L' J' a' \Gamma'
\gamma') \; c (\alpha' S' L' J' a' \Gamma' ; f) \eqno (9)
$$
in terms of $O(3) \supset G$ symmetry adapted state vectors.

The only good quantum numbers for the initial and final state vectors are
$\Gamma \gamma$ and $\Gamma' \gamma'$, respectively. Although, the state
vectors (7) and (9) are developed in a weak-field basis, it is to be noted that
the intensity calculation to be conducted in what follows is valid for any 
(weak-, intermediate- or strong-field) coupling scheme. 
The expansion coefficients
$c (\alpha S L J a \Gamma ; i)$ and $c (\alpha' S' L' J' a' \Gamma' ; f)$ can
be obtained by optimizing an Hamiltonian involving at least Coulomb, spin-orbit
and crystal-field interactions ; the introduction of more sophisticated
interactions in the Hamiltonian may be useful to take covalency effects into
account [3]. Alternatively, the expansion coefficients in (7) and (9) can be
considered as free parameters entering in the phenomenological intensity
parameters to be introduced below.

\noindent {\bf Transition matrix element}

By using a quasi-closure approximation, it can be shown that the transition 
matrix element $M_{i \to f}$ between the state vectors 
$\vert i \Gamma  \gamma )$ and 
$\vert f \Gamma' \gamma')$ turns out to be given by 
$$
M_{i \to f} \; \equiv \; M_{i (\Gamma \gamma) \to f (\Gamma' \gamma')} 
\; = \; (f \Gamma' \gamma' \vert H_{eff} \vert i \Gamma \gamma) \eqno (10)
$$
where $H_{eff}$ is an effective operator [12-15]. This operator may 
be written as [7]
$$
H_{eff} \; = \; \sum_{k_Sk_Lk} 
  \; C \left[ \left( k_S k_L \right) k \right] 
  \; \left( \left\{ {\cal E}_1 \, {\cal E}_2 \right\} ^{(k)} . \, 
  \; {\bf W}^{(k_Sk_L)k} \right) \eqno (11)
$$
In equation (11), ${\bf W}^{(k_Sk_L)k}$ is an electronic double tensor of 
spin rank $k_S$, orbital rank $k_L$ and total rank $k$. The information on the
polarization of the two photons is contained in the tensor product
$\left\{ {\cal E}_1 \, {\cal E}_2 \right\} ^{(k)}$ 
of rank $k = 0$, 1 or 2. The right-hand side
of (11) is a development in terms of scalar products $(\,.\,)$ with expansion
coefficients $C\left[\left(k_S k_L\right)k\right]$. These coefficients depend
on the ground configuration $n \ell^N$ and on the configurations 
$n  \ell ^{N-1} n' \ell'$ and/or 
$n' \ell'^{4 \ell' + 1} n \ell^{N +1}$, with $(-)^{\ell + \ell'} = - 1$, 
from which the virtual states $v$ arise. 

Only the contributions 
$(k_S = 0, k_L = 1, k = 1)$ and 
$(k_S = 0, k_L = 2, k = 2)$ correspond to the
standard theory originally developed by Axe [12]. The other contributions 
$(k_S \ne 0, k_L, k)$, which may include 
$(k_S = 1, k_L = 1, k = 0)$ and 
$(k_S = 1, k_L = 1, k = 2)$, 
correspond either to mechanisms introduced by various authors 
[13-15] or to phenomenological contributions introduced in the spirit 
of Ref.~[7]. The contributions $(k_S =   0, k_L = k, k)$ and 
                               $(k_S \ne 0, k_L    , k)$ are often referred 
to as second-order and third-order mechanisms, respectively. It is in principle
possible to find an expression for the parameters 
$C \left[ \left( k_S k_L \right) k \right] $. 
For example, for the contribution $\left( k_S = 0, k_L = k, k \right)$ we have
$$
C[(0k)k] = - \sqrt 2 \; e^2 \; \sum_{n' \ell '} \; 
\left[ (-)^k (\Delta'_1)^{-1} + (\Delta'_2)^{-1} \right] \;
(n \ell \vert r \vert n' \ell')^2 \; 
\left( \ell \Vert C^{(1)} \Vert \ell' \right)^2 \;
\left\{
\matrix{
1&k&1\cr
\ell&\ell'&\ell
}
\right\}
\eqno (12)
$$
where, according to the quasi-closure approximation, 
$\Delta'$ replaces $\Delta$ of equation (1) with
$$
\Delta'_\lambda \; = \; \hbar \omega_\lambda \; - \; E (n' \ell') \qquad 
\hbox {for} \qquad \lambda = 1,2
\eqno (13)
$$
(Most of the other symbols in equation (12), and in the rest of this 
paper, have their usual significance.) 
Among the various 
contributions $(k_S \ne 0, k_L, k)$, the contribution 
$(k_S = 1, k_L = 1, k = 0)$ arises from the spin-orbit 
interaction within the configuration $n \ell^{N-1} n' \ell'$ as 
was shown for lanthanide ions [13,14].

The transition matrix element (10) is easily calculated by means of 
Wigner-Racah calculus for the chain $O(3) \supset G$. As a result, we have 
(see Ref.~[7])
$$
\eqalign{
M_{i(\Gamma \gamma) \to f (\Gamma' \gamma')} = & \sum_{\alpha' S' L' J' a'} \; 
                                                 \sum_{\alpha  S  L  J  a } 
\; c (\alpha' S' L' J' a' \Gamma' ; f)^* 
\; c (\alpha  S  L  J  a  \Gamma  ; i)\cr
& \sum_{k_S k_L k} \; (-)^{k_S + k_L - k} \; 
C \left[ \left( k_S k_L \right) k\right] \; 
\left( n \ell^N \alpha  S  L  J \Vert W^{(k_S k_L)k} \Vert 
       n \ell^N \alpha' S' L' J' \right) ^*\cr
& \sum_{a'' \Gamma'' \gamma''} \, f
\pmatrix{
J&J'&k\cr
a \Gamma \gamma& a' \Gamma' \gamma'& a'' \Gamma'' \gamma''\cr
}^* \,  
\left\{ {\cal E}_1 \, {\cal E}_2 \right\} ^{(k)}_{a'' \Gamma'' \gamma''}
\cr
}
\eqno (14)
$$
where the $f$ symbol denotes an $O(3) \supset G$ symmetry adapted coupling
coefficient defined by
$$
\eqalign{
f
\pmatrix{
J&J'&k\cr
a \Gamma \gamma & a' \Gamma' \gamma' & a'' \Gamma'' \gamma''\cr
}
\; = \; \sum_{M M' q} \;
&(-)^{J-M}
\pmatrix{
J&k&J'\cr
-M&q&M'\cr
} \cr
&(J  M  | J  a   \Gamma   \gamma )^* \; 
 (J' M' | J' a'  \Gamma'  \gamma')   \; 
 (k  q  |  k a'' \Gamma'' \gamma'') 
} \eqno (15) 
$$
Equation (14) immediately follows by developing (10) with the 
help of (7), (9) and (11).

\noindent {\bf Intensity formula} 

The quantity of interest for a comparison between theory and experiment is the
intensity $S_{i(\Gamma) \to f(\Gamma')}$ of the two-photon transition between
the initial state $i$ and the final state $f$. This intensity is given by
$$
S_{  \Gamma  \to   \Gamma' } \; \equiv \; 
S_{i(\Gamma) \to f(\Gamma')} \; = \; \sum_{\gamma \gamma'} \; 
\left\vert M_{i(\Gamma \gamma) \to f(\Gamma' \gamma')} \right\vert ^2 
\eqno (16)
$$
By introducing (14) into (16), we get an expression involving
$$
X \; = \; \sum_{\gamma \gamma'} \; f
\pmatrix{
J&J'&k\cr
a \Gamma \gamma & a' \Gamma' \gamma' & r \Gamma'' \gamma''\cr
}^*
\; f
\pmatrix{
\bar J & \bar J' & \ell \cr
\bar a \Gamma \gamma & \bar a' \Gamma' \gamma'
& s \bar \Gamma'' \bar \gamma''\cr
}
\eqno (17)
$$
i.e., a sum over $\gamma$ and $\gamma'$ of the product of two 
particular $f$ coefficients. This sum can be calculated to be
$$
X = ([J] [\bar J])^{-1/2} \; [\Gamma'']^{-1} \; [\Gamma] 
\; \delta (\bar \Gamma '' , \Gamma '')
\; \delta (\bar \gamma '' , \gamma '') \; 
\sum_{\beta} \; 
(     J'      a' \Gamma' + k    r \Gamma'' \vert      J      a \beta \Gamma) \; 
(\bar J' \bar a' \Gamma' + \ell s \Gamma'' \vert \bar J \bar a \beta \Gamma) ^* 
\eqno (18)
$$
To derive the sum rule (18), it is sufficient to apply twice the 
factorization property [16]
$$
\eqalign{
f
\pmatrix{
j_1&j_2&k\cr
a_1 \Gamma_1 \gamma_1 & a_2 \Gamma_2 \gamma_2 & a \Gamma \gamma\cr
}
& \; = \; (-)^{2k} \; [j_1]^{-1/2}\cr
& \sum_\beta \; (j_2a_2 \Gamma_2 + k a \Gamma \vert j_1
a_1 \beta \Gamma_1)^* \; 
(\Gamma_2 \Gamma \gamma_2 \gamma \vert 
 \Gamma_2 \Gamma \beta \Gamma_1 \gamma_1)^*\cr
}
\eqno(19)
$$
for the $f$ symbol and once the orthonormality-completeness property 
[16]
$$
\eqalign{
\sum_{\gamma_1 \gamma} \; 
  (\Gamma_1 \Gamma _2 \gamma_1 \gamma _2 \vert 
   \Gamma_1 \Gamma _2 \beta  \Gamma \gamma)^* \;
& (\Gamma_1 \Gamma'_2 \gamma_1 \gamma'_2 \vert 
   \Gamma_1 \Gamma'_2 \beta' \Gamma \gamma)   \;  = \cr 
& \Delta(\Gamma \vert \Gamma_1 \otimes \Gamma_2) \; 
\delta (\Gamma'_2, \Gamma_2) \; 
\delta (\gamma'_2, \gamma_2) \; \delta (\beta', \beta) \; 
[\Gamma_2]^{-1} \; [\Gamma]\cr 
} \eqno (20)
$$
for the Clebsch-Gordan coefficients of $G$. 
The $( \; + \; \vert \; )$ coefficients in equations (18) and (19) 
stand for isoscalar factors of the chain $O(3) \supset G$. In 
(18) and (19), the labels of type $\beta$ are internal multiplicity 
labels to be used
for those Kronecker products which are not multiplicity-free. The introduction
of (18) into (16) leads to the compact expression
$$
S_{\Gamma \to \Gamma'} \; = \; \sum_{k \ell} \; \sum_{r s} \; \sum_{\Gamma''} \;
I [k \ell r s \Gamma'' ; \Gamma \Gamma'] \; \sum_{\gamma''} \; \left\{ 
{\cal E}_1 \, {\cal E}_2 \right\} ^{(k)}    _{r \Gamma'' \gamma''} \; 
\left( \left\{ 
{\cal E}_1 \, {\cal E}_2 \right\} ^{(\ell)} _{s \Gamma'' \gamma''}\right)^* 
\eqno (21)
$$
In equation (21), the parameter $I[ \cdots ]$ reads
$$
\eqalign{
I [k \ell r s \Gamma'' ; \Gamma \Gamma'] \; = \; [\Gamma'']^{-1} \; [\Gamma] 
\; & \sum_{J'a'} \; 
     \sum_{J a } \; 
\sum_{\bar J' \bar a'} \; 
\sum_{\bar J  \bar a}\cr
& Y_k    (     J'      a' \Gamma' ,      J      a \Gamma) \; 
  Y_\ell (\bar J' \bar a' \Gamma' , \bar J \bar a \Gamma)^*\cr
&\sum_\beta \; (J' a' \Gamma' + k r \Gamma'' \vert J a \beta \Gamma) \; 
(\bar J' \bar a' \Gamma'+\ell s \Gamma'' \vert \bar J \bar a \beta \Gamma)^*\cr
}
\eqno(22)$$
where $Y_k$ is defined by
$$
\eqalign{
Y_k (J' a' \Gamma', J a \Gamma) \; = \; & [J]^{-1/2} \; \sum_{\alpha' S' L'} \; 
\sum_{\alpha S L} \; \sum_{k_S k_L}\cr
& c (\alpha' S' L' J' a' \Gamma' ; f)^* \; c (\alpha S L J a \Gamma ; i) 
\; C [(k_S k_L) k]\cr
& (-)^{k_S + k_L - k} \; 
(n \ell^N \alpha S L J \Vert W^{(k_Sk_L)k} \Vert n \ell^N \alpha' S' L' J')^*\cr
}
\eqno (23)
$$
and $Y_\ell$ by a relation similar to (23).

\noindent {\bf Properties and rules}

The $I$ parameters in (21) can be calculated in an {\it ab initio} way or 
can be considered as phenomenological parameters. In both approaches, the 
following properties and rules are of central importance.

{\bf Property 1}. In the general case, we have the (hermitean) property
$$
I[\ell k s r \Gamma'' ; \Gamma \Gamma']^* \; = \; 
I[k \ell r s \Gamma'' ; \Gamma \Gamma'] \eqno (24)
$$
which ensures that $S_{\Gamma \to \Gamma'}$ is a real quantity.

{\bf Property 2}. In the case where the group $G$ is multiplicity-free, 
we have the factorization formula
$$
I[k \ell r s \Gamma'' ; \Gamma \Gamma'] \; = \; 
\chi [k    r \Gamma'' ; \Gamma \Gamma'] \; 
\chi [\ell s \Gamma'' ; \Gamma \Gamma']^*
\eqno (25)
$$
where the function $\chi$ is defined through
$$
\chi [k    r \Gamma'' ; \Gamma \Gamma'] \; = \; 
[\Gamma'']^{-1/2} \; [\Gamma]^{1/2} \; \sum_{J'a'} \; \sum_{Ja} \; 
Y_k(J' a' \Gamma', J a \Gamma) \; 
(J' a' \Gamma' + k r \Gamma'' \vert J a \Gamma)
\eqno (26)
$$
(In a less restrictive sense, equation (25) is valid when the Kronecker 
product $\Gamma'^* \otimes \Gamma$, of the complex conjugate IRC of 
$\Gamma'$ by the IRC $\Gamma$, is multiplicity-free.)

The number of independent parameters $I[ \cdots ]$ in the expansion 
(21) can be {\it a priori} determined from the two following selection 
rules used in conjunction with Properties 1 and 2.

{\bf Rule 1}. In order to have $S_{\Gamma \to \Gamma'} \ne 0$, it is
necessary that
$$
\Gamma'' \subset \Gamma'^* \otimes \Gamma 
\eqno (27)
$$
and
$$
\Gamma'' \subset (k_g) \qquad \quad \Gamma'' \subset (\ell_g) 
\eqno (28)
$$
where $(k_g)$ and $(\ell_g)$ are {\it gerade} IRC's of the group 
$O(3)$ associated to the integers $k$ and $\ell$, respectively.

{\bf Rule 2}. The sum over $k$ and $\ell$ in the intensity 
formula (21) is partially controlled by the selection rule
$$
\eqalign{
 {\cal E}_1 \ne {\cal E}_2 \qquad 
& k, \ell = 1,2     \quad  \hbox {for} \quad k_S =   0 \cr
& k, \ell = 0, 1, 2 \quad  \hbox {for} \quad k_S =   0 \quad \hbox {and} 
                                       \quad k_S \ne 0 \cr
}
\eqno (29)
$$
or 
$$
\eqalign{
 {\cal E}_1 =   {\cal E}_2 \qquad  
& k, \ell = 2       \quad  \hbox {for} \quad k_S   = 0 \cr
& k, \ell = 0, 2    \quad  \hbox {for} \quad k_S   = 0 \quad \hbox {and} 
                                       \quad k_S \ne 0 \cr
}
\eqno (30)
$$
according to as the two photons have different or the same polarization. 
(Note that the situation ${\cal {E}}_1 = {\cal {E}}_2$ surely occurs for 
identical photons but may also occur for non-identical photons.)

\noindent {\bf Discussion}

For most of the cases of interest, there is no summation on $r$ and $s$, two
branching multiplicity labels of type $a$, in the intensity formula 
(21). (In other words, the frequency of $\Gamma''$ in $(k_g)$ and $(\ell_g)$ 
is rarely greater than 1.) The group-theoretical selection rules 
(27) and (28) impose strong limitations
on the summation over $\Gamma''$ in (21) once $\Gamma$ and $\Gamma'$ are
fixed and the range of values of $k$ and $\ell$ is chosen.

The number of independent intensity parameters $I[ \cdots ]$ in 
the formula (21) is determined by~: (i) the nature of the photons, cf.~Rule 2~;
(ii) the group $G$, cf.~Rule 1~; (iii) the symmetry property (24), cf.~Property 
1~; (iv) the use of $k_S = 0$ (second-order mechanism) or $k_S = 0$ and
$k_S \not = 0$ (second- plus third-order mechanisms), cf.~Rule 2~; 
(v) the kind of the (weak-, intermediate- or 
strong-field) coupling used for the state vectors, cf.~equations (22) and (23).

Points (i)-(iii) depend on external physical conditions. On the other hand,
points (iv) and (v) are model-dependent. In particular, in the case where the
$J$-mixing, cf.~point (v), can be neglected, a situation of interest for
lanthanide ions, the summations on $k$ and $\ell$ in (21) are further reduced
by the triangular rule $|J-J'| \leq k, \ell \leq J+J'$, where $J$ and $J'$ are
the total angular quantum numbers for the initial and final states,
respectively. Similar restrictions apply to $k_S$ and $k_L$ in 
(23) if the $S$- and $L$-mixing are neglected.

The computation, via equations (22) and (23), of the $I$ parameters generally 
is a difficult task. Therefore, they may be considered, at least in a
first step, as phenomenological parameters. In this respect, equations 
(22) and (23) should serve as a guide for reducing the number of $I$ parameters.

Once the number of independent parameters $I[ \cdots ]$ in the intensity
formula (21) has been determined, we can obtain the polarization dependence 
of the intensity strength 
$S_{\Gamma \to \Gamma'}$ by calculating the tensor products 
$\left\{ {\cal E}_1 \, {\cal E}_2 \right\} ^{(K)} _{a'' \Gamma'' \gamma''}$ 
(with $K = k, \ell$ and $a'' = r, s$) occurring in equation (21). 
For this purpose, we use the development 
$$
\left\{ {\cal E}_1 \, {\cal E}_2 \right\} ^{(K)} _{a''\Gamma'' \gamma''} 
\; = \;  \sum^K_{Q = - K} \;
\left\{ {\cal E}_1 \, {\cal E}_2 \right\} ^{(K)} _Q \; 
(K Q \vert K a'' \Gamma'' \gamma'')
\eqno (31)
$$
in terms of the spherical components 
$\left\{ {\cal E}_1 \, {\cal E}_2 \right\} ^{(K)} _Q$,
the coefficients in the development (31) being reduction coefficients 
for the chain $O(3) \supset G$. Then, we use in turn the development
$$
\left\{ {\cal E}_1 \, {\cal E}_2 \right\} ^{(K)} _Q \; = \; 
(-)^{K - Q} \; [K]^{1/2} \; \sum^1_{x = - 1} \; \sum^1_{y = -1} \; 
\pmatrix{
1&K&1\cr
x&-Q&y\cr
}
\; ({\cal E}_1)_x 
\; ({\cal E}_2)_y 
\eqno (32)
$$
in terms of the spherical components $({\cal E}_\lambda)_q$ defined 
by (4) or (5) for circular or linear polarization, respectively.

\noindent {\bf Illustration}

As an illustrative example, we consider the case of the configuration $nd^8$ in
cubical symmetry with $G \equiv O$. Let us examine the intra-configurational 
two-photon absorption transitions from the initial state $i~=~^3A_2(T_2)$ to
the final states taken as the first excited states $f~=~^3T_2(E,T_1,T_2,A_2)$. 
Therefore,
we have $\Gamma = T_2$ for the initial state and $\Gamma ' = A_2,E,T_1,T_2$ for
the various final states. Furthermore, 
there is no sum on the multiplicity labels 
$r$ and $s$ in the intensity formula (21). Let us begin
with non-identical photons. Then, the possible values of $k$ and $\ell$ in 
(21) are $0,1,2$. Since the restriction $SO(3) \to O$ yields 
$$
(0) = A_1 \qquad \quad (1) = T_1 \qquad \quad (2) = E \oplus T_2
\eqno (33)
$$
we have $k = \ell$ in (21). Consequently, the intensity parameters 
$I[ \cdots ]$ assume the form
$$
I[kk \Gamma'' ; \Gamma \Gamma'] \; = \; 
  \vert \chi [k \Gamma'' ; \Gamma \Gamma'] \vert ^2 \qquad 
  {\hbox {with}} \qquad \Gamma'' = A_1, T_1, E, T_2
\eqno (34)
$$
since the group $O$ is multiplicity-free. 
More precisely, we are left with 10 independent parameters for non-identical
photons~; we shall take the following normalization~:
$$
\eqalign{
  a_{T_2T_2}(0A_1) \; &= \; {1\over 3} \; I[00A_1;T_2T_2]\cr
  a_{T_2T_2}(2T_2) \; &= \; {1\over 4} \; I[22T_2;T_2T_2]\cr
  a_{T_2T_1}(2E)   \; &= \; {1\over 6} \; I[22E;T_2T_1]\cr
  a_{T_2T_1}(1T_1) \; &= \; {1\over 4} \; I[11T_1;T_2T_1]\cr
  a_{T_2E}(1T_1)   \; &= \; {1\over 4} \; I[11T_1;T_2E]\cr
}\qquad \quad
\eqalign{
  a_{T_2T_2}(2E)   \; &= \; {1\over 6} \; I[22E;T_2T_2]\cr
  a_{T_2T_2}(1T_1) \; &= \; {1\over 4} \; I[11T_1;T_2T_2]\cr
  a_{T_2T_1}(2T_2) \; &= \; {1\over 4} \; I[22T_2;T_2T_1]\cr
  a_{T_2E}(2T_2)   \; &= \; {1\over 4} \; I[22T_2;T_2E]\cr
  a_{T_2A_2}(1T_1) \; &= \; {1\over 4} \; I[11T_1;T_2A_2]\cr
}\eqno (35)
$$
As a result, the intensity strengths are given by
$$
\eqalign{
  S_{T_2 \rightarrow T_2}
 &= 3 \; a_{T_2T_2} (2E) \; + \; 2 \; a_{T_2T_2} (2T_2) \cr 
  S_{T_2 \rightarrow T_1} 
 &= 3 \; a_{T_2T_1} (2E) \; + \; 2 \; a_{T_2T_1} (2T_2) \cr
  S_{T_2 \rightarrow E} 
 &= 2 \; a_{T_2E} (2T_2) \cr 
  S_{T_2 \rightarrow A_2}
 &= 0 \cr
 }\eqno (36)
$$
for circular polarization (the two photons 
having the same circular polarization) and by 
$$
\eqalign{
  S_{T_2 \rightarrow T_2}
 &= a_{T_2T_2} (0A_1) \; \varpi_1 \; + \; 
    a_{T_2T_2} (2E)   \; \varpi_2 \; + \; 
    a_{T_2T_2} (2T_2) \; \varpi_3 \; + \; 
    a_{T_2T_2} (1T_1) \; \varpi_4\cr
  S_{T_2 \rightarrow T_1} 
 &= a_{T_2T_1} (2E)   \; \varpi_2 \; + \; 
    a_{T_2T_1} (2T_2) \; \varpi_3 \; + \; 
    a_{T_2T_1} (1T_1) \; \varpi_4\cr
  S_{T_2 \rightarrow E} 
 &= a_{T_2E} (2T_2) \; \varpi_3 \; + \; 
    a_{T_2E} (1T_1) \; \varpi_4\cr 
  S_{T_2 \rightarrow A_2}
 &= a_{T_2A_2} (1T_1) \; \varpi_4\cr
 }\eqno (37)
$$
for linear polarization. The angular functions $\varpi_i$ ($i = 1,2,3,4$) 
in equation (37) read
$$
\matrix{
  \varpi_1 = 
 &[       \cos \theta_1 \; \cos \theta_2
 &+ \quad \sin \theta_1 \; \sin \theta_2 \;
  \cos (\varphi_1 - \varphi_2) ]^2 \hfill \cr
  \varpi_2 = 
 &[2      \cos \theta_1 \; \cos \theta_2  
 &- \quad \sin \theta_1 \; \sin \theta_2 \;
  \cos (\varphi_1 - \varphi_2) ]^2 \hfill \cr
&&+ \quad 3 \sin^2 \theta_1 \; \sin^2 \theta_2 \; \cos^2 (\varphi_1 + \varphi_2)
  \hfill \cr
  \varpi_3 = 
 &2         \sin^2 \theta_1 \; \cos^2 \theta_2 
 &+ \quad 2 \cos^2 \theta_1 \; \sin^2 \theta_2 \hfill \cr
 &&+ \quad \sin 2 \theta_1 \; \sin 2 \theta_2 \; \cos (\varphi_1 - \varphi_2) + 
       2 \sin^2 \theta_1 \; \sin^2 \theta_2 \; \sin^2 (\varphi_1 + \varphi_2)
  \hfill \cr
  \varpi_4 = 
 &2         \sin^2 \theta_1 \; \cos^2 \theta_2 
 &+ \quad 2 \cos^2 \theta_1 \; \sin^2 \theta_2 \hfill \cr
 &&- \quad \sin 2 \theta_1 \; \sin 2 \theta_2 \; \cos (\varphi_1 - \varphi_2) 
     + 2 \sin^2 \theta_1 \; \sin^2 \theta_2 \; \sin^2 (\varphi_1 - \varphi_2)
 \hfill \cr
}\eqno (38)
$$
We now continue with the particular case where the two photons are identical.
In this case, we get $\varpi_4 \equiv 0$ so that the number of independent
parameters in equations (36) and (37) is reduced from 10 to 6~: the parameter
$a_{T_2T_2}(0A_1)$ describes third-order mechanisms while the parameters
$a_{T_2T_2}(2E)$, $a_{T_2T_2}(2T_2)$, $a_{T_2T_1}(2E)$, 
$a_{T_2T_1}(2T_2)$ and
$a_{T_2E}(2T_2)$ may be thought to mainly describe second-order mechanisms.

Similar results hold for the other intra-configurational two-photon transitions
of $nd^8$ in $O$. For example, let us consider the transition between 
the inital state $i~=~^3A_2(T_2)$ with $\Gamma = T_2$ and 
the final  state $f~=~^3T_1(A_1)$ with $\Gamma' = A_1$. By putting 
$$
a_{T_2A_1}(2T_2) \; = \; {1\over 4} \; I[22T_2 ; T_2A_1]
\eqno (39)
$$
we obtain (for identical or non-identical photons)
$$
S_{T_2 \rightarrow A_1} \; = \; a_{T_2A_1} (2T_2) \; \varpi_3 
\qquad {\hbox {or}} \qquad 2 \; a_{T_2A_1} (2T_2)
\eqno (40)
$$
according to as the polarization is linear or circular. 

Indeed, all the two-photon transitions arising from an initial state of
symmetry $T_2$ are given by formulas of the type (36), (37) and (39). 
The formulas (36) and (37) generalize to the case of non-identical 
photons the formulas for identical photons derived in Ref.~[18] 
in order to explain the experimental results of Ref.~[17] 
concerning Ni$^{2+}$ in MgO.

\noindent {\bf III. INTER-CONFIGURATIONAL TWO-PHOTON TRANSITIONS}

\noindent {\bf Sketch of the theory}

We now consider two-photon transitions between Stark levels 
arising from the configurations $n \ell^N$ and 
$n \ell^{N-1} n' \ell'$ of opposite parities 
($(-)^{\ell + \ell'} = -1$). For the sake of simplicity, we 
deal here with identical photons. The initial and final state 
vectors are (respectively) taken in the form 
$$
\vert i \Gamma \gamma) \; \equiv \; \vert n \ell^N  i \Gamma \gamma) 
\; = \; 
\sum_{\alpha S L J a} \; \vert n \ell^N \alpha S L J a \Gamma \gamma) 
\; c(n \ell^N ; \alpha S L J a \Gamma ; i) 
\eqno (41)
$$
and
$$
\eqalign{
\vert f \Gamma' \gamma') \; \equiv \; 
\vert n \ell^{N-1} n' \ell' f \Gamma' \gamma') 
\; = \; \sum_{\alpha' S' L' J' a'} \; 
 & \vert n \ell^{N-1} n' \ell'   \alpha' S' L' J' a' \Gamma' \gamma') \; \cr 
 &    c (n \ell^{N-1} n' \ell' ; \alpha' S' L' J' a' \Gamma' ; f) \cr 
} \eqno (42)
$$
to be compared with equations (7) and (9).

It is clear that the transition matrix element
$$
M_{i(\Gamma \gamma) \to f(\Gamma' \gamma')} \; = \; \sum_v \; 
{{1}\over {\Delta}} \; 
\left( f \Gamma' \gamma'  \vert \vec D. \, \vec {\cal E} \vert 
      v \Gamma_v \gamma_v \right) 
\left(v \Gamma_v \gamma_v \vert \vec D. \, \vec {\cal E} 
\vert i \Gamma \gamma \right) 
\eqno (43)
$$
is identically zero. In order to obtain 
$M_{i(\Gamma \gamma) \to f(\Gamma' \gamma')} \ne 0$, it is 
necessary to pollute the state vectors (41) and (42) with state 
vectors of the type $\vert n \ell^{N-1} n' \ell' x' \Gamma' \gamma')$ 
and                 $\vert n \ell^{N}            x  \Gamma  \gamma )$, 
respectively. This may be 
achieved by using first-order perturbation theory where the 
polluting agent is the crystal-field potential $H_3$ of odd order, 
which is static or dynamic according to as the group $G$ does not or does 
have a center of inversion. We thus produce state vectors 
noted $\vert n \ell^ N             i \Gamma  \gamma >$ and 
      $\vert n \ell^{N-1} n' \ell' f \Gamma' \gamma'>$ from which we 
can calculate a non-vanishing transition matrix element 
$$
M_{i(\Gamma \gamma) \to f(\Gamma' \gamma')} \; = \; \sum_v \; 
{{1}\over {\Delta}} \; <f \Gamma' \gamma' 
                     \vert \vec D. \, \vec {\cal E} \vert v \Gamma_v \gamma_v>
<v \Gamma_v \gamma_v \vert \vec D. \, \vec {\cal E} \vert i \Gamma \gamma> 
\eqno (44)
$$
Then, we apply a quasi-closure approximation both for the 
initial, intermediate (virtual), and final state 
vectors and the transition matrix element. This approximation can be 
summarized by 
$$
E(n' \ell') - E(n \ell) \; = \; 2 \, \hbar \, \omega
\eqno (45)
$$ 
We thus obtain a closed form formula for 
$M_{i(\Gamma \gamma) \to f(\Gamma' \gamma')}$. 

At this stage, it should be mentioned that the so-obtained 
formula is equivalent to that we would obtain, within the same 
approximation (45), by using third order mechanisms described 
by
$$
\eqalign{
M_{i(\Gamma \gamma) \to f(\Gamma' \gamma')} & = \cr
& \sum_{v_1 v_2} \; {{1}\over {\Delta(v_1)}} \; {{1}\over {\Delta(v_2)}} 
\; ( f \Gamma' \gamma' 
                             \vert \vec D. \, \vec {\cal E} 
\vert v_1 \Gamma_1 \gamma_1 ) ( v_1 \Gamma_1 \gamma_1
\vert \vec D. \, \vec {\cal E} 
\vert v_2 \Gamma_2 \gamma_2 ) ( v_2 \Gamma_2 \gamma_2
\vert H_3 \vert i \Gamma \gamma ) \cr 
+ & \sum_{v_1 v_2} \; {{1}\over {\Delta(v_1)}} \; {{1}\over {\Delta(v_2)}} 
\; ( f \Gamma' \gamma' 
\vert \vec D. \, \vec {\cal E} 
\vert v_1 \Gamma_1 \gamma_1 ) ( v_1 \Gamma_1 \gamma_1
\vert H_3  
\vert v_2 \Gamma_2 \gamma_2 ) ( v_2 \Gamma_2 \gamma_2
\vert \vec D. \, \vec {\cal E} \vert i \Gamma \gamma ) \cr
+ & \sum_{v_1 v_2} \; {{1}\over {\Delta(v_1)}} \; {{1}\over {\Delta(v_2)}} 
\; ( f \Gamma' \gamma' \vert H_3 
\vert v_1 \Gamma_1 \gamma_1 ) ( v_1 \Gamma_1 \gamma_1 
\vert \vec D. \, \vec {\cal E} 
\vert v_2 \Gamma_2 \gamma_2 ) ( v_2 \Gamma_2 \gamma_2
\vert \vec D. \, \vec {\cal E} \vert i \Gamma \gamma ) \cr 
} \eqno (46)
$$
where the initial, intermediate (virtual) and final state 
vectors are non-polluted state vectors.

By following the same line of reasoning as in the case of 
intra-configurational transitions, we are left with the 
intensity formula
$$
\eqalign{
S_{\Gamma \to \Gamma'} \; = \; {\hbox {Re}} \; [
& \sum_{k, \ell = 0,2} \; \sum_{r s} \; \sum_{\Gamma''} \;
I_1 [k \ell r s \Gamma'' ; \Gamma \Gamma'] \; \sum_{\gamma''} \; \left\{ 
{\cal E} \, {\cal E} \right\} ^{(k)}    _{r \Gamma'' \gamma''} \; 
\left( \left\{ 
{\cal E} \, {\cal E} \right\} ^{(\ell)} _{s \Gamma'' \gamma''}\right)^* 
\cr
& + \sum_{ k = 0,2} \; \sum_{r s} \; \sum_{\Gamma''} \;
I_2 [k    2 r s \Gamma'' ; \Gamma \Gamma'] \; \sum_{\gamma''} \; \left\{ 
        {\cal E} \, {\cal E} \right\} ^{(k)} _{r \Gamma'' \gamma''} \; 
\left\{ {\cal E} \, {\cal E} \right\} ^{(2)} _{s \Gamma'' \gamma''} 
] \cr 
}
\eqno (47)
$$
which parallels the formula (21).

\noindent {\bf Illustration}

Let us consider the case of the configuration $4f$ in tetragonal symmetry 
with $G \equiv C_{4v}$ and examine the two-photon transitions
between the Stark levels of the shells $4f$ and $5d$
(i.e., $n \ell \equiv 4f$, $N \equiv 1$, $n' \ell' \equiv 5d$). 
There are four possible transitions since the initial and final 
states may have the symmetries $\Gamma_6$ and $\Gamma_7$. For a 
linear polarization, the application of the intensity formula 
(47) leads to 
$$
\eqalign{
S_{\Gamma_6 \to \Gamma_6} \; & = \; 
a  + b \, \pi_1 + c \, \pi_1^2 + d \, \pi_2 + e \, \pi_3 \cr 
S_{\Gamma_7 \to \Gamma_7} \; & = \;
a' + b'\, \pi_1 + c'\, \pi_1^2 + d'\, \pi_2 + e'\, \pi_3 \cr 
S_{\Gamma_6 \to \Gamma_7} \; & = \;
f \, \pi_2 + g \, \pi_3 + h \, \pi_4 + i \, \pi_5 \cr
S_{\Gamma_7 \to \Gamma_6} \; &= \;
f'\, \pi_2 + g'\, \pi_3 + h'\, \pi_4 + i'\, \pi_5 \cr
}\eqno (48)
$$
where the angular functions $\pi_i$ ($i=1,2,3,4,5$) are defined by
$$
\pi_1 = 3 \; \cos^2 \theta \; - \; 1        \quad 
\pi_2 = \sin^2 2 \theta                     \quad 
\pi_3 = \sin^2 2 \theta \; \cos   2 \varphi \quad 
\pi_4 = \sin^4   \theta \; \cos^2 2 \varphi \quad 
\pi_5 = \sin^4   \theta \; \sin^2 2 \varphi 
\eqno (49)
$$
The various parameters $a, \cdots, i$ and $a', \cdots, i'$ 
are simple functions of the intensity parameters $I_1[ \cdots ]$ and 
$I_2[ \cdots ]$ occurring in (47).

\noindent {\bf IV. CLOSING REMARKS}

We have shown how $O(3) \supset G$ symmetry adaptation allows 
to derive intensity formulas for intra- and 
inter-configurational two-photon transitions for ions in 
molecular or solid-state environments. In particular, the 
number of independent parameters required for describing the 
polarization dependence of the transitions is determined by an 
ensemble of properties and rules which combine symmetry and 
physical considerations. The main results of this paper are 
equations (21) and (47) for intra- and inter-configurational 
transitions, respectively. The case of intra-configurational 
transitions has been treated in detail. The case of 
inter-configurational symmetry shall be developed again in 
forthcoming papers and in the thesis by one of us [11].

The reader should consult Refs.~[18,19] where symmetry adaptation 
techniques, in the spirit of the present paper, have been applied 
to rare-earth and transition-metal ions in various symmetries.

Thanks are due to J.C.~G\^acon and B.~Jacquier for 
discussions, on the occasion of a series of seminars during the 
academic year 1990-91, and to J.~Sztucki for correspondence. 
The senior author (M.~K.) is grateful to the 
organizing committee of the ``V Workshop on Symmetry Methods in 
Physics'' for inviting him in Moscow and Obninks. 

\noindent {\bf REFERENCES}

\baselineskip 0.5 true cm

\itemitem{[1]} D.T. Sviridov and Yu.F. Smirnov, Theory of 
Optical Spectra of Transition-Metal Ions (Nauka, Moscow, 1977).

\itemitem{[2]} Tang Au-chin, Sun Chia-chung, Kiang Yuan-sun, 
Deng Zung-hau, Liu Jo-chuang, Chang Chain-er, Yan Guo-sen, Goo 
Zien and Tai Shu-san, Theoretical Method of the Ligand Field 
Theory (Science Press, Beijing, 1979).
 
\itemitem{[3]} M. Kibler and G. Grenet, Studies in 
Crystal-Field Theory (Report LYCEN/8656, IPNL, Lyon, 1986).

\itemitem{[4]} M. Inoue and Y. Toyozawa, J. Phys. Soc. Japan 
20 (1965) 363. 

\itemitem{[5]} T.R. Bader and A. Gold, Phys. Rev. 171 (1968) 997.

\itemitem{[6]} P.A. Apanasevich, R.I. Gintoft, V.S. Korolkov, A.G.
Makhanek and G.A. Skripko, Phys. Status Solidi (b) 58 (1973) 
745~; A.G. Makhanek and G.A. Skripko, Phys. Status Solidi (a) 53 (1979) 
243~; L.A. Yuguryan, Preprints N$^\circ$ 232 and 233, Inst. Fiz.
Akad. Nauk BSSR, Minsk (1980).

\itemitem{[7]} M. Kibler and J.C. G\^acon, Croat. Chem. Acta 62
(1989) 783. 

\itemitem{[8]} M. Kibler, in~: Symmetry and Structural Properties 
of Condensed Matter, Eds.  W. Florek, T. Lulek and M. Mucha 
(World, Singapore, 1991).  

\itemitem{[9]} R. Loudon, The Quantum Theory of Light 
(Clarendon, Oxford, 1973).

\itemitem{[10]} C. Cohen-Tannoudji, J. Dupont-Roc et G. 
Grynberg, Processus d'interaction entre photons et atomes 
(InterEditions et Editions du CNRS, Paris, 1988).

\itemitem{[11]} M. Daoud, th\`ese de Doctorat (Universit\'e 
Lyon-1, in preparation). 

\itemitem{[12]} J.D. Axe, Jr., Phys. Rev. 136 (1964) A42.

\itemitem{[13]} B.R. Judd and D.R. Pooler, J. Phys. C : Solid State Phys.
15 (1982) 591.

\itemitem{[14]} M.C. Downer and A. Bivas, Phys. Rev. B~28 (1983) 
3677~; M.C. Downer, G.W. Burdick and D.K. Sardar, J. Chem. 
Phys. 89 (1988) 1787~; M.C. Downer, in~: Laser Spectroscopy of Solids II, 
Ed. W.M. Yen (Springer, Heidelberg, 1989).

\itemitem{[15]} J. Sztucki and W. Str\c ek, Phys. Rev. B~34 (1986) 
3120~; Chem. Phys. 143 (1990) 347.

\itemitem{[16]} M. Kibler, C.R. Acad. Sc. (Paris) B 268 (1969) 
1221~; M.R. Kibler and P.A.M. Guichon, Int. J. Quantum 
Chem. 10 (1976) 87~; M.R. Kibler, in~: Recent Advances in Group 
Theory and Their Application to Spectroscopy, Ed. J.C. Donini 
(Plenum Press, N.Y., 1979)~; Int. J. Quantum Chem. 23 (1983) 115.

\itemitem{[17]} C. Campochiaro, D.S. McClure, P. Rabinowitz 
and S. Dougal, Phys. Rev. B~43 (1991) 14. 

\itemitem{[18]} J. Sztucki, M. Daoud and M. Kibler, Phys. Rev. 
B (to be submitted for publication).

\itemitem{[19]} J.C. G\^acon, J.F. Marcerou, M. Bouazaoui, B. Jacquier
and M. Kibler, Phys. Rev. B~40 (1989) 2070~; J.C. G\^acon, B. Jacquier, 
J.F. Marcerou, M. Bouazaoui and M. Kibler, J. Lumin. 45 (1990) 
162~; J.C. G\^acon, M. Bouazaoui, B. Jacquier, M. Kibler, L.A. 
Boatner and M.M. Abraham, Eur. J. Solid State Inorg. Chem. 28 (1991) 113. 

\bye